\documentclass{amsart}[11pt]
\usepackage{graphicx}

\begin{document}

\title{Application of deep quantum neural networks to finance}

\author{Takayuki Sakuma}
\thanks{Faculty of Economics, Soka University. e-mail: tsakuma@soka.ac.jp} 
\keywords{option pricing, neural networks, quantum machine learning, differential machine learning.}
\date{\today}

\begin{abstract}
The recent development of quantum computing gives us an opportunity to explore its potential applications to many fields, with the field of 
finance being no exception. In this paper, we apply the deep quantum neural network proposed by Beer et al. (2020) and discuss such potential in the context of 
simple experiments such as learning implied volatilities and option prices. Furthermore, Greeks such as delta and gamma, which are important measures in risk management, can be computed analytically with the neural network, 
and our numerical experiments show that the deep quantum neural network is a promising technique for solving such numerical problems arising in finance efficiently.\end{abstract}

\maketitle


\section{Introduction}
Recently, the rapid development of quantum computing has been eye-opening, and it is not an overstatement to say that the use of quantum computers in daily life will be realized in the near future.
Therefore the great potential of quantum computing in the financial industry has been actively investigated. For example, the Monte Carlo method is a widely used tool for computing derivative prices in the financial industry currently.
Stamatopoulos et al. (2020) applied a quantum algorithm called amplitude estimation proposed by Brassard et al. (2002) to derivative pricing and achieved a quadratic speed-up, which is more efficient relative to the Monte Carlo method. Application to the valuation of Value at Risk (VaR) and Conditional Value at Risk (C-VaR) was also considered in Woerner and Egger (2019).
Furthermore, the quantum algorithm has also been applied by Ramos-Calderer et al. (2019) for an asset probability distribution mapped to unary representation. 
Another approach applying quantum computing was proposed by Fontanela et al. (2019), in which the original partial differential equation satisfied by the option price is transformed to a Schrödinger equation in imaginary time and the equation is then solved numerically using a hybrid algorithm, with both classical and quantum computing. 

In this paper, we discuss the potential of applying the deep quantum neural network (DQNN) proposed by Beer et al. (2020) to the field of finance.  
The DQNN is a natural generalization of the classical deep neural network and we discuss this potential in the context of simple experiments such as learning implied volatilities and option prices. 
Greeks such as delta and gamma, which are important measures in risk management, can be computed analytically using the DQNN once it has learned the option prices. 
Our numerical experiments show that the deep quantum neural network are a promising technique for solving numerical problems arising in finance efficiently.
\\
\\
\section{Basics of quantum computing: qubit}
In classical computers, the basic component of information is called a bit, which is expressed by 0 or 1 deterministically. 
On the other hand, quantum computers use two basis quantum states denoted as $|0 \rangle$ and $|1 \rangle$, analogously to a classical bit,  to express their basis component of information, which is called a qubit and determined probabilistically.
Mathematically, the state of a qubit $|\phi\rangle$ is expressed as 
\\
\begin{equation*}
  \label{eqn:ioeq0}
  |\phi\rangle= \alpha |0 \rangle+\beta |1 \rangle,
\end{equation*}
\\
where $\alpha$ and $\beta$ are complex number and the squares $\alpha^2$ and $\beta^2$ are the probabilities of observing 
states $|0 \rangle$ and $|1 \rangle$, respectively. Quantum computing utilizes this superposition property to achieve much higher 
computational efficiency than classical computers. For example, quantum supremacy was achieved in 2019; Arute et al. (2019) 
reported that Google succeed in computing a problem in only 200 seconds using a quantum computer that would take 200 million 
years to solve with a classical supercomputer. At the same time, classical deep learning is a currently 
active research topic in finance and has been shown to be efficient for solving high-dimensional problems, including those 
arising in finance (for example, Gnoatto et al. (2020)). Therefore the use of the DQNN, fusing idea of deep learning to quantum computing, is a natural next step for discovering highly efficient methods for solving the numerical problems arising in finance.
\\
\section{Deep quantum neural networks}
Here, we give a brief explanation of the DQNN proposed by Beer et al. (2020). Suppose some quantum states 
$|\phi^{in}\rangle$ and $|\phi^{out}\rangle$ satisfy the following relation: 
\\
\begin{equation}
\label{eqn:ioeq1}
|\phi^{out}\rangle= U |\phi^{in}\rangle,
\end{equation}
\\
where $U$ is an unknown unitary operator which is to be learned by the DQNN using $N$ training pairs ($|\phi_n^{in}\rangle,|\phi_n^{out}\rangle$)
for $n=1,..,N$.
\\
\\
The architecture of the DQNN is analogous to that of a classical deep neural network: the DQNN is composed of an input layer, 
an output layer, and $L$ hidden layers. The main issue is how to achieve a feedforward propagation 
of information in the DQNN because the clone theorem states that the quantum state of a qubit cannot be copied to another qubit. 
\\
\\
The first key property of the DQNN is that the density matrix of output quantum state $\rho_{D}^{out}$ can be expressed as 
\\
\begin{equation}
  \label{eqn:ioeq2222}
  \rho_{D}^{out} =\varepsilon^{out}(\varepsilon^{L}(...(\varepsilon^2(\varepsilon^1(\rho^{in})))...)),
\end{equation}
\\
where
\\
\begin{equation}
  \label{eqn:ioeq23}
  \varepsilon^{l}(X^{l-1})\equiv {tr}_{l-1}(\prod_{j=m_l}^{l}U_j^l(X^{l-1}\otimes |0...0\rangle_l \langle 0...0)\prod_{j=1}^{m_l} \mathcal{U}_j^{l \dag}),
\end{equation}
\\
$m_l$ is the number of unitary matrices acting on the ($l-1$)-th layer, and $U_j^l$ is the $j$-th unitary matrix acting on the ($l-1$)-th layer. Equation (\ref{eqn:ioeq23}) indicates that, at the $l$-th layer, the quantum state expressed as density matrix $X^{l-1}$ is tensorized with qubits $|0...0 \rangle$, unitaries $\prod_{j=m_l}^{l}U_j^l$ are 
applied to the tensorized quantum state, and finally the resulting quantum state is traced out over ($l-1$)-th qubits. This expression indicates that the information from the input flows from layer to layer can be regarded as equivalent to an expression in 
the backpropagation algorithm utilized in conventional neural networks. Furthermore, (\ref{eqn:ioeq23}) can be expressed using the Kraus operator $A_\alpha$:
\\
\begin{equation}
  \label{eqn:ioeq223}
  \varepsilon^{l}(X^{l-1})\equiv\sum_\alpha A_\alpha X^{l-1} A_\alpha^\dagger.
\end{equation}
\\
Similar to a classical deep quantum network, we want to estimate $U$ from the training pairs by minimizing some cost function. We define cost function $C$ as
\\
\begin{equation}
\label{eqn:ioeq3}
C\equiv \frac{1}{N}\sum_{n=1}^{N}\langle\phi_n^{out}|\rho_{D,n}^{out}|\phi_n^{out}\rangle,
\end{equation}
\\
which varies between 0 and 1. Its form reflects the idea of fidelity, which represents how close the output state $\rho_{D,n}^{out}$ 
of (\ref{eqn:ioeq2222}) with input state $|\phi_n^{in}\rangle$ is to the training state $|\phi_n^{out}\rangle$: $C$ takes a value of 1 if the output state is identical to the training state.
\\
\\
For parameter updating using a quantum computer, Taylor approximation is proposed in Beer et al. (2020). And also, parameter updating using a classical computer was proposed as follows:
\\
\begin{equation}
\label{eqn:ioeq4}
U \rightarrow e^{i K \epsilon}U,
\end{equation}
\\
where $K$ represents the parameterized matrix chosen such that $C$ increases the most rapidly. That is,
$K$ is determined such that it maximizes $\Delta C$:
\\
\begin{equation}
\label{eqn:ioeq5}
\Delta C = \frac{\epsilon}{N} \sum_{n=1}^{N} \sum_{l=1}^{L+1} tr((\sigma_n^l) (\Delta \varepsilon^{l}) (\rho_n^{l-1})),
\end{equation}
\\
where $\epsilon$ is the step size, \begin{math} \rho_n^l =\varepsilon^l(...(\varepsilon^2(\varepsilon^1(\rho_n^{in}))...) \end{math},
\begin{math}\sigma_n^l=\mathcal{F}^{l+1}(...\mathcal{F}^{out}(|\phi_n^{out}\rangle \langle\phi_n^{out}|)...) \end{math},
and $\mathcal{F}$ is the adjoint channel of $\varepsilon$ given by
\\
\begin{equation}
  \label{eqn:ioeq51}
  \mathcal{F}^{l}(X^{l-1})\equiv\sum_\alpha A_\alpha^\dagger X^{l-1} A_\alpha.
\end{equation}
\\
Then, it is derived in Beer et al. (2020) that the elements of $K$ are represented as follows:
\\
\begin{equation}
\label{eqn:ioeq6}
K_j^l=\eta \frac{2^{m_{l-1}}}{N} \sum_{n=1}^{N}tr_{rest} M_j^l,
\end{equation}
\\
where
\\
\begin{equation}
\label{eqn:ioeq7}
M_j^l=[\prod_{\alpha=j}^{1}U^l_{\alpha}(\rho_n^{l-1}\otimes|0...{0\rangle}_l\langle0...0|) 
\prod_{\alpha=1}^{j}{U^l_{\alpha}}^{\dag},
\prod_{\alpha=j+1}^{m_l}{U^l_{\alpha}}^{\dag}(\mathbb{I}_{l-1}\otimes \sigma_n^l) 
\prod_{\alpha=m_l}^{j+1}U^l_{\alpha}],
\end{equation}
\\
$\eta$ is the learning rate and $rest$ in (\ref{eqn:ioeq6}) means that $M_j^l$ is traced out over all the qubits independent of unitary $U^l_{\alpha}$. 
\\
\\
The second remarkable property of the DQNN is that, in order to evaluate $K_j^l$, only the output of the previous 
layer, $\rho_n^{l-1}$, which is obtained using feedforward, and the state of the following layer, $\sigma_n^l$, are needed. 
This helps reduce the memory requirement for executing the DQNN. The trade-off here is that multiple evaluations
of the network are needed; however, this is not a hard task for NISQs. 
\\
\\
Throughout this paper, we use the following notation:
\\
\begin{equation}
  \label{eqn:ioeq888}
  P(s,t,u) \equiv \frac{1}{1+\exp(-(\frac{s}{t})^u)}\cdot \frac{\pi}{2}.
\end{equation}
\\
\section{Application 1: Learning implied volatilities} 
As a first application, we consider the learning of implied volatilities in the option market: 
the input data constitute the strike ${K_1,K_2,...,K_N}$ and the output data constitute the corresponding implied volatility ${\sigma_{K_1},\sigma_{K_2},...,\sigma_{K_N}}$.
The number of qubits that we use in the input layer and the output layer is  one, and in order to use these financial data in the DQNN, we first need to convert them to quantum states as follows for $n=1,2,...,N$:
\\
\begin{equation}
\label{eqn:ioeq8}
|\phi_n^{in}\rangle= \cos (P(F,K_n,\beta))|0\rangle+\sin (P(F,K_n,\beta))|1\rangle,
\end{equation}
\begin{equation}
\label{eqn:ioeq9}
|\phi_n^{out}\rangle= \cos (P(\sigma_{K_n},K_n,\gamma))|0\rangle+\sin (P(\sigma_{K_n},K_n,\gamma))|1\rangle,
\end{equation}
\\
where $F$ is an underlying and $\beta,\gamma$ are fixed parameters. The sigmoid function is used to convert market data to numbers 
between 0 and 1, and to fuse them into Bloch sphere representations. For simplicity, time to maturity is fixed and, 
as in Ruf and Wang (2020), we can extend it to estimating volatility surfaces by setting $\sigma_{K_n} \sqrt{\tau}$
instead of $\sigma_{K_n}$, where $\tau$ is time to maturity.
\\
\\
The density matrix of training data $|\phi_n^{out}\rangle$ is $|\phi_n^{out}\rangle \langle\phi_n^{out}|$ and we can extract learned implied volatilities $\sigma_{D,K_n}$ from the (1,1) element $X$ of matrix $\rho_{D,n}^{out}$:
\\
\begin{equation}
\label{eqn:ioeq10}
\sigma_{D,K_n}=K_n(-\log(\frac{\pi}{2 \arccos(\sqrt{X})}-1))^\frac{1}{\gamma}.
\end{equation}
\\
If we implement the DQNN using a quantum computer, we can calculate $\cos (P(\sigma_{K_n},K_n,\gamma))$ 
by computing the probability of observing state $|0\rangle$ of the qubit in the output layer.
\\
\\
As numerical examples, we use market data taken from Antonov et al. (2015) and publicly available MATLAB code 
\footnote{The code is available at https://github.com/qigitphannover/DeepQuantumNeuralNetworks.} for DQNN simulation.
The network consists of one input qubit, one output qubit, and one hidden layer with two qubits ($L$=1, $m_1=2$). 
The seeds for generating random initial states of $U_j^l$ are set as 0. 
The learning rate $\lambda$ is 1, $\epsilon$ is 0.1, and the number of iterations to update $U$ is 800. Tables 1 to 3 summarize the comparison results 
for the output and the training data with different $\beta$ and $\gamma$. 
Table 4 shows the results for three qubits in one hidden layer ($L$=1, $m_1=3$), and a slight improvement in the performance can 
be seen in this case. Increasing the number of qubits in a hidden layer does not result in a substantial improvement because 
this simple numerical experiment only uses one input layer and one output layer. 
\\
\\
Additionally, an arbitrage-free condition, which the implied volatility surface must satisfy, is not taken into account and
we can add the following arbitrage-free conditions to the DQNN as in Itkin (2020):
\\
\begin{equation*}
\label{eqn:ioeq32}
\frac{dV}{dK}<0, \frac{d^2 V}{d K^2}>0,\frac{dV}{d \tau}>0.
\end{equation*}
\\
Itkin (2020) imposed penalties in the MSE loss function if these conditions are not satisfied. However, this requires the efficient estimation of Greeks, which is our next application of DQNN.
\\
\section{Application 2: Learning option prices and computing Greeks}
Suppose $V$ is the value of a financial instrument and $x$ is an underlying asset that affects $V$. The estimation of not only $V$ but also the differentials called Greeks such as $\frac{dV}{dx}$ and $\frac{d^2V}{dx^2}$ is 
an important task for financial risk management. Differential machine learning (DML) proposed by 
Huge and Savine (2020) has received great attention in the financial industry because DML estimates these differentials efficiently 
using a twin network scheme. After using a conventional feedforward neural network for predicting $V$, DML applies 
backpropagation via automatic adjoint differentiation (AAD) to compute the differential of the output $V$ with respect to the input $x$. Weights are updated to minimize a cost function consisting of a weighted summation of 
the prediction errors of $V$ and the differentials. Another aspect of DML is that it can approximate the shape of $V$ with respect to $x$ via the differentials, which improves the estimation accuracy of $V$ greatly.
\\
\\
The success of DML in the financial industry motivated us to consider a second application: learning option prices and Greeks using the DQNN.
Given $x$, $V$, and strike $K$, we consider the following input and output:
\\
\begin{equation*}
\label{eqn:ioeq10}
|\phi^{in}\rangle=\phi_{1}^{in}|0\rangle+\phi_{2}^{in}|1\rangle,
|\phi^{out}\rangle=\phi_{1}^{out}|0\rangle+\phi_{2}^{out}|1\rangle,
\end{equation*}
where
\begin{equation*} 
\label{eqn:ioeq10tt}
\phi_{1}^{in}=\cos (P(x,K,\beta)), \phi_{2}^{in}=\sin (P(x,K,\beta))
\end{equation*}
and
\begin{equation*}
  \label{eqn:ioeq10ttt}
\phi_{1}^{out}=\cos (P(V,K,\gamma)),\phi_{2}^{out}=\sin (P(V,K,\gamma)).
\end{equation*}
\\
Then density matrices $\rho^{in}$ and $\rho^{out}$ are given as
\\
\begin{equation*}
  \label{eqn:ioeq11t}
  \rho^{in}
   = \left(
    \begin{array}{cc}
      (\phi_{1}^{in})^2 & \phi_{1}^{in}\phi_{2}^{in}\\
      \phi_{1}^{in}\phi_{2}^{in} & (\phi_{2}^{in})^2
    \end{array}
    \right),
    \rho^{out}    
    = \left(
     \begin{array}{cc}
      (\phi_{1}^{out})^2 & \phi_{1}^{out}\phi_{2}^{out}\\
       \phi_{1}^{out}\phi_{2}^{out} & (\phi_{2}^{out})^2
     \end{array}
     \right)
\end{equation*}
\\
and similar to the first application, the DQNN gives
\\
\begin{equation}
  \label{eqn:ioe1211t}
  \rho_{D}^{out} = {tr}_{in,hid} ({U}(\rho^{in}\otimes|0...0{\rangle}_{hid,out} \langle0...0|){U}^{\dag}).
\end{equation}
\\
The right-hand side of (\ref{eqn:ioe1211t}) means that $\rho^{in} $ is tensorized with qubits $|0...0\rangle$ in hidden layers and the output 
layer, unitaries $U$ are applied to the tensorized quantum state, and finally the resulting quantum state is traced out over qubits in input layer and hidden layers.
\\
\\
Furthermore, given (\ref{eqn:ioe1211t}), it holds that
\\
\begin{equation}
  \label{eqn:ioeq12ttttt}
  \frac{d\rho_{D}^{out}}{d x} = {tr}_{in,hid} ({U}(\frac{d\rho^{in}}{d x}\otimes|0...0{\rangle}_{hid,out} \langle0...0|){U}^{\dag}).
  \end{equation}
\\
Equation (\ref{eqn:ioeq12ttttt}) implies that given the DQNN and $\frac{d\rho^{in}}{d x}$, we can compute $\frac{d\rho_{D}^{out}}{d x}$. Therefore, the DQNN is highly effective for learning differentials.
\\
\\
Now We have
\\
\begin{equation*}
  \label{eqn:ioeq13ttttt}
  \frac{d \rho^{in}}{d x}
   = \left(
    \begin{array}{cc}
      \frac{d ({\phi_{1}^{in}})^2}{d x} & \frac{d (\phi_{1}^{in}\phi_{2}^{in})}{d x} \\
      \frac{d (\phi_{1}^{in}\phi_{2}^{in})}{d x} & \frac{d ({\phi_{2}^{in}})^2}{d x} 
    \end{array}
    \right)
    = \left(
      \begin{array}{cc}
        2\phi_{1}^{in}\frac{d {\phi_{1}^{in}}}{d x} & \phi_{1}^{in}\frac{d {\phi_{2}^{in}}}{d x}+\phi_{2}^{in}\frac{d {\phi_{1}^{in}}}{d x} \\
        \phi_{1}^{in}\frac{d {\phi_{2}^{in}}}{d x}+\phi_{2}^{in}\frac{d {\phi_{1}^{in}}}{d x}  & 2\phi_{2}^{in}\frac{d {\phi_{2}^{in}}}{d x}.
      \end{array}
      \right),
\end{equation*}
\\
where 
\\
\begin{equation*} 
\frac{d \phi_{1}^{in}}{d x}=-k_{x} \cdot \phi_{2}^{in}, \frac{d \phi_{2}^{in}}{d x}=k_{x}\cdot \phi_{1}^{in}
\end{equation*}
\\
and
\\
\begin{equation*}
  \label{eqn:ioeq13tttttttttt}
  k_{x}=\frac{\pi}{2} \cdot \frac{\beta}{K} \cdot {(\frac{x}{K})}^{\beta-1}\frac{1}{1+\exp(-{(\frac{x}{K})}^{\beta})}[1-\frac{1}{1+\exp(-{(\frac{x}{K})}^{\beta})}].
\end{equation*}
\\
Thus, we have 
\\
\begin{equation*}
  \label{eqn:ioeq13tttttttt}
  \frac{d \rho^{in}}{d x} = k_{x} \cdot M_{\rho, x}, 
\end{equation*}
\\
where
\\
\begin{equation*}
  \label{eqn:ioeq13tttttttttt}
  M_{\rho, x}
    \equiv \left(
      \begin{array}{cc}
        -\sin (2 P(x,K,\beta)) & \cos (2 P(x,K,\beta)) \\
        \cos (2 P(x,K,\beta))  &  \sin (2 P(x,K,\beta))
      \end{array}
      \right).
\end{equation*}
\\
Similarly,
\\
\begin{equation}
  \label{eqn:ioeq155t}
  \frac{d \rho^{out}}{d x}
    = \frac{d V}{d x} \cdot k_{V} \cdot M_{\rho, V},
\end{equation}
\\
where
\\
\begin{equation*}
  \label{eqn:ioeq15tttttt}
k_{V}=\frac{\pi}{2} \cdot \frac{\gamma}{K} (\frac{V}{K})^{\gamma-1} \cdot\frac{1}{1+\exp(-{(\frac{V}{K})}^{\gamma}}[1-\frac{1}{1+\exp(-{(\frac{V}{K})}^{\gamma}}]
\end{equation*}
\\
and
\\
\begin{equation*}
  \label{eqn:ioeq15tt}
  M_{\rho, V}
    = \left(
      \begin{array}{cc}
         -\sin (2 P(V,K,\gamma))  & \cos (2 P(V,K,\gamma)) \\
         \cos (2 P(V,K,\gamma))  & \sin (2 P(V,K,\gamma))
      \end{array}
      \right).
\end{equation*}
\\
If the DQNN is implemented in a quantum computer, the main issue is that $\frac{d\rho^{in}}{d x}$ and $\frac{d\rho^{out}}{d x}$ are not quantum states; the traces are zero. 
The traces must be one and the eigenvalues must be positive real numbers and we therefore modify them as follows:
\\
\begin{equation*}
{\frac{d \rho^{in,m}}{d x}}\equiv \frac{\mu_{in}}{2}\frac{d \rho^{in}}{d x}+\frac{I}{2} = \frac{1}{2}(r_{in} M_{\rho, x}+I), r_{in}=\mu_{in} \cdot k_{x},
\end{equation*}
\begin{equation*}
{\frac{d \rho^{out,m}}{d x}}\equiv  \frac{\mu_{out}}{2}\frac{d \rho^{out}}{d x}+\frac{I}{2} = \frac{1}{2}(r_{out} M_{\rho, V}+I), r_{out}=\mu_{out} \cdot \frac{d V}{d x}\cdot k_{V},
\end{equation*}
\\
where $I$ is the identity matrix and $\mu_{in},\mu_{out}$ are scaling parameters. 
Note $\frac{d\rho^{in,m}}{d x}$ and $\frac{d\rho^{out,m}}{d x}$ are quantum states since the traces are equal to one and the eigenvalues are positive real as long as
\\
\begin{equation*}
\label{eqn:condition11}
0 \le r_{in} \le \frac{1}{2},0 \le r_{out} \le \frac{1}{2}.
\end{equation*}
\\
Then the following output from the DQNN is also a quantum state:
\\
\begin{equation*}
  \label{eqn:ioeq12tttttttt}
  \frac{d\rho_{D}^{out,m}}{d x} \equiv {tr}_{in,hid} ({U}(\frac{d\rho^{in,m}}{d x}\otimes|0...0{\rangle}_{hid,out} \langle0...0|){U}^{\dag}).
\end{equation*}
\\
We can define $\frac{d^2 \rho^{in,m}}{d {x}^2}$ and $\frac{d^2 \rho^{out,m}}{d {x}^2}$ similarly.
Then, the cost function to be maximized is
\\
\begin{equation*}
\label{eqn:ioeq17}
C_d\equiv \frac{1}{N}\sum_{n=1}^{N}\langle\phi_n^{out}|\rho_{D,n}^{out}|\phi_n^{out}\rangle + 
\nu_d \frac{1}{N}\sum_{n=1}^{N} {f_d}({\frac{d \rho_{n}^{out,m}}{d x}}, {\frac{d \rho_{D,n}^{out,m}}{d x}})+
\nu_g \frac{1}{N}\sum_{n=1}^{N} {f_d}({\frac{d^2 \rho_{n}^{out,m}}{d {x}^2}}, {\frac{d^2 \rho_{D,n}^{out,m}}{d {x}^2}}),
\end{equation*}
\\
where $f_d(S,T)$ represents the fidelity of density matrices $S$ and $T$, and $\nu_d$ and $\nu_g$ control the relative influence of the differential fidelity on the cost function. 
The unitary layers are updated as
\\
\begin{equation*}
  \label{eqn:ioeq18}
  U \rightarrow e^{i K_d \epsilon}U,
  \end{equation*}
\\
where $K_d$ represents the parameterized matrix that is chosen such that $C_d$ increases the most rapidly. 
A similar argument to that in the supplementary information of Beer et al. (2020) gives the elements of $K_d$ as follows:
\\
\begin{equation*}
  \label{eqn:ioeq19}
  K_{d,j}^l=\eta \frac{2^{m_{l-1}}}{N} (\sum_{n=1}^{N}tr_{rest} M_j^l+\nu_d \sum_{n=1}^{N}tr_{rest} M_{d,j}^l+\nu_g \sum_{n=1}^{N}tr_{rest} M_{g,j}^l),
\end{equation*}
\\
where
\\
  \begin{equation*}
  \label{eqn:ioeq20}
  M_{d,j}^l=[\prod_{\alpha=j}^{1}U^l_{\alpha}({\rho_{d,n}}^{l-1}\otimes|0...{0\rangle}_{hidden,out}\langle0...0|) 
  \prod_{\alpha=1}^{j}{U^l_{\alpha}}^{\dag},
  \prod_{\alpha=j+1}^{m_l}{U^l_{\alpha}}^{\dag}(\mathbb{I}_{l-1}\otimes  {\sigma_{d,n}}^l) 
  \prod_{\alpha=m_l}^{j+1}U^l_{\alpha}],
  \end{equation*}
\\
  \begin{math} {\rho_{d,n}}^l =\varepsilon^l(...(\varepsilon^2(\varepsilon^1({\frac{d \rho_{n}^{in,m}}{d x}}))...) \end{math},
  \begin{math}{\sigma_{d,n}}^l=\mathcal{F}^{l+1}(...\mathcal{F}^{out}({\frac{d \rho_{n}^{out,m}}{d x}})...) \end{math},
\\
  \begin{equation*}
    \label{eqn:ioeq20}
    M_{g,j}^l=[\prod_{\alpha=j}^{1}U^l_{\alpha}({\rho_{g,n}}^{l-1}\otimes|0...{0\rangle}_{hidden,out}\langle0...0|) 
    \prod_{\alpha=1}^{j}{U^l_{\alpha}}^{\dag},
    \prod_{\alpha=j+1}^{m_l}{U^l_{\alpha}}^{\dag}(\mathbb{I}_{l-1}\otimes  {\sigma_{g,n}}^l) 
    \prod_{\alpha=m_l}^{j+1}U^l_{\alpha}],
    \end{equation*}
    \begin{math} {\rho_{g,n}}^l =\varepsilon^l(...(\varepsilon^2(\varepsilon^1({\frac{d^2 \rho_{n}^{in,m}}{d {x}^2}}))...) \end{math},
    \begin{math}{\sigma_{g,n}}^l=\mathcal{F}^{l+1}(...\mathcal{F}^{out}({\frac{d^2 \rho_{n}^{out,m}}{d {x}^2}})...) \end{math}.
\\
\\
Finally, the predicted $V$ is given as 
\\
\begin{equation*}
\label{eqn:ioeq21}
K(-\log(\frac{\pi}{2 \arccos(\sqrt{X})}-1))^\frac{1}{\gamma},
\end{equation*}
\\
where $X$ is the (1,1) element of matrix $\rho_{D}^{out}$. If we implement the DQNN using a quantum computer, we can 
calculate $\cos (P(V_n,K,\gamma))$ by computing an approximation of the probability of observing $|0\rangle$.
\\
\\
Similarly, the predicted delta, $\frac{d V}{d x}$, is given as 
\\
\begin{equation*}
\label{eqn:ioeq22}
-\frac{Y-Z}{ \frac{\mu_{out}}{2} \cdot k_{V_{p}} \cdot \sin (\frac{1}{1+\exp(-{(\frac{V_{p}}{K})}^{\gamma})}\cdot \pi)},
\end{equation*}
\\
where $Y$ is the (1,1) element of matrix ${\frac{\rho_{D}^{out,m}}{d x}}$, $Z$ is the (1,1) element of 
\begin{math} {tr}_{in,hid} ({U}(\frac{1}{2}I\otimes|0...0{\rangle}_{hid,out} \langle0...0|){U}^{\dag}) \end{math},
$V_{p}$ is the option price predicted by the DQNN, and
\\
\begin{equation*}
  \label{eqn:23}
k_{V_{p}}=\frac{\pi}{2} \cdot \frac{\gamma}{K} (\frac{V_{p}}{K})^{\gamma-1} \cdot\frac{1}{1+\exp(-{(\frac{V_{p}}{K})}^{\gamma}}[1-\frac{1}{1+\exp(-{(\frac{V_{p}}{K})}^{\gamma}}].
\end{equation*}
\\
\\
We can compute the predicted gamma $\frac{d^2 V}{d x^2}$ similarly. 
\\
\\
Note that ${\frac{d \rho^{out,m}}{d x}}$ can be expressed as
\\
\begin{equation*}
  {\frac{d \rho^{out,m}}{d x}} =\frac{1}{2}(r_{out} M_{\rho, V_{n}}+I)=\frac{1}{2}I+r_{out}\cos(2P(V_n,K, \gamma))X-r_{out}\sin(2P(V_n,K, \gamma)) Z,
\end{equation*}
where $X$ and $Z$ are Pauli matrices
\\
\begin{equation*}
  X
    = \left(
      \begin{array}{cc}
         0  & 1 \\
         1  & 0
      \end{array}
      \right),
      Z
      = \left(
        \begin{array}{cc}
           1  & 0 \\
           0  & -1
        \end{array}
        \right).
\end{equation*}
\\
This implies that the Bloch vector of ${\frac{d \rho^{out,m}}{d x}}$ is 
\\
\begin{equation*}
(r_{out}\cos(2P(V_n,K, \gamma)),0,-r_{out}\sin((2P(V_n,K, \gamma)))
\end{equation*}
\\
and if we implement the DQNN using a quantum computer, we can compute $-r_{out}\sin((2P(V_n,K, \gamma))$ and the predicted delta by measuring the expected value of the qubit in the output layer in the $Z$-basis. 
\\
\\
In a numerical simulation, we use publicly available MATLAB code as well and we set $\mu_{in}=\mu_{out}=2$. We use seven training sets of spot price $x$, call price $V$, delta $\frac{d V}{d x}$,  and gamma $\frac{d^2 V}{d x^2}$ under the Black--Scholes model with strike $K=100$, $r=0.0$, $T=0.25$, and volatility $=0.15$. 
Here, $\beta$ is set as $5.0$ and $\gamma$ is set as $0.5$.  The network consists of one input layer with one qubit, one output layer with one qubit, 
and one hidden layer with three qubits ($L$=1, $m_1=3$). Tables 5, 6, and 7 summarize the comparison results for the output and 
the training data with different spot prices $S$. Higher order differentials depend on the value of lower order differentials, so the accuracy of lower
order differentials is especially important; an increase of $\nu_d$ or $\nu_f$ does not affect the accuracy of delta or gamma, respectively,  if the difference between 
the training price and predicted price is not small. 
\\
\section{Conclusion}
The possibility of applying the DQNN proposed by Beer et al. (2020) to finance is investigated. Applications to learning implied volatilities and option prices are explored, and the obtained simple numerical results demonstrate that the DQNN can be considered to be a promising candidate 
for developing highly powerful methods in finance, since Beer et al. (2020) has already shown that the DQNN exhibit some degree of robustness to noise, which is a major issue in noisy intermediate-scale quantum devices (NISQs). As a next step, further careful investigation will be carried out by working with large datasets. Also, how the DQNN can be implemented efficiently using a quantum computer is an important research topic. Although Beer et al. (2020) proposed one appropriate approach, in order to implement it, we would need a quantum computer with high capacity in the sense of it needing to be able to implement SWAP gates and partial trace operations efficiently.
Lastly, quantum neural networks (QNNs) have been pointed out to have the ``barren plateau" problem by McClean et al. (2018), that is, that gradients vanish during parameter optimization. This issue arises in QNNs but not in conventional neural networks, and the DQNN is no exception, as shown in Sharma (2020). Therefore, the DQNN needs to be constructed carefully to avoid this issue, but remains a very promising tool for use in finance. 

\newpage

\begin{table}[htb]
  \begin{center}
    \caption{Experimental results: $F=0.56\%$, $\beta=1.0$, $\gamma=1.0$, $C=0.9984$}
    \begin{tabular}{|c||c|c|c|} \hline
       Strike ($\%$) & Training $\sigma$ (bps) & Output & Diff\\ \hline \hline
       0.06 & 23.5 & 17.2 & 6.3  \\
       0.31 & 44.7 & 53.5 & -8.8 \\
       0.56 & 59.3 & 62.6 & -3.3 \\
       0.81 & 71.7 & 71.0 & 0.7 \\
       1.06 & 83.0 & 79.7 & 3.3 \\
       1.56 & 103.5 & 97.6 &5.9 \\
       2.56 & 140.5 & 134.2 &6.2 \\ \hline
     \end{tabular}
    \end{center}
\end{table}   

\begin{table}[htb]
    \begin{center}
      \caption{Experimental results: $F=0.56\%$, $\beta=0.5$, $\gamma=0.5$, $C=0.9998$}
      \begin{tabular}{|c||c|c|c|} \hline
       Strike ($\%$) & Training $\sigma$ (bps) & Output & Diff\\ \hline \hline
       0.06 & 23.5 & 21.9 & 1.6  \\
       0.31 & 44.7 & 48.3 & -3.6 \\
       0.56 & 59.3 & 61.3 & -2.0 \\
       0.81 & 71.7 & 72.2 & -0.5 \\
       1.06 & 83.0 & 82.2 & 0.8 \\
       1.56 & 103.5 & 100.7 &2.8 \\
       2.56 & 140.5 & 134.7 &5.7 \\ \hline
      \end{tabular}
    \end{center}
\end{table}    

\begin{table}[htb]
    \begin{center}
      \caption{Experimental results: $F=0.56\%$, $\beta=0.25$, $\gamma=0.25$, $C=0.9997$}    
      \begin{tabular}{|c||c|c|c|} \hline
       Strike ($\%$) & Training $\sigma$ (bps) & Output & Diff\\ \hline \hline
       0.06 & 23.5 & 21.4 & 2.1  \\
       0.31 & 44.7 & 46.7 & -2.0 \\
       0.56 & 59.3 & 61.5 & -2.2 \\
       0.81 & 71.7 & 73.4 & -1.7 \\
       1.06 & 83.0 & 83.8 & -0.8 \\
       1.56 & 103.5 & 101.9 &1.6 \\
       2.56 & 140.5 & 132.6 &7.8 \\ \hline
      \end{tabular}
    \end{center}
\end{table}     

\begin{table}[htb]
    \begin{center}
      \caption{Experimental results: $F=0.56\%$, $\beta=0.25$, $\gamma=0.25$, $m_1=3$, $C=0.9999$}   
      \begin{tabular}{|c||c|c|c|} \hline
      Strike ($\%$) & Training $\sigma$ (bps) & Output & Diff\\ \hline \hline
      0.06 & 23.5 & 21.1 & 2.4  \\
      0.31 & 44.7 & 46.1 & -1.4 \\
      0.56 & 59.3 & 61.4 & -2.1 \\
      0.81 & 71.7 & 73.8 & -2.1 \\
      1.06 & 83.0 & 84.8 & -1.8 \\
      1.56 & 103.5 & 104.2 &-0.7 \\
      2.56 & 140.5 & 137.5 &2.9 \\ \hline
      \end{tabular}
      \end{center}
  \end{table}
    
  \newpage

\begin{table}[htb]
  \begin{center}
    \caption{Predicted call price}
     \begin{tabular}{|c||c|c|c|c|} \hline
       x & Training & $\nu_d=\nu_g=0.0$ & $\nu_d=\nu_g=0.01$ & $\nu_d=0.05,\nu_g=0.02$ \\ \hline \hline
       93	&0.640	&0.689	&0.690	&0.694	\\
       95	&1.073	&1.141	&1.142	&1.146	\\
       97	&1.686	&1.748	&1.749	&1.753	\\
      100	&2.991	&2.990	&2.992	&3.001	\\
      103	&4.769	&4.665	&4.671	&4.692	\\
      105	&6.193	&6.024	&6.035	&6.073	\\
      107	&7.776	&7.563	&7.580	&7.644	\\ \hline
     \end{tabular}
    \end{center}
\end{table}
 
\begin{table}[htb]
    \begin{center}
      \caption{Predicted delta}
     \begin{tabular}{|c||c|c|c|c|} \hline
       x & Training & $\nu_d=\nu_g=0.0$ & $\nu_d=\nu_g=0.01$ & $\nu_d=0.05,\nu_g=0.02$ \\ \hline \hline
       93	&0.176	&0.191	&0.191	&0.191	\\
       95	&0.259	&0.263	&0.263	&0.263	\\
       97	&0.356	&0.346	&0.346	&0.346	\\
      100	&0.515	&0.485	&0.486	&0.488	\\
      103	&0.667	&0.632	&0.634	&0.640	\\
      105	&0.754	&0.726	&0.729	&0.739	\\
      107	&0.826	&0.810	&0.815	&0.829	\\ \hline
      \end{tabular}
  \end{center}
\end{table}
  
\begin{table}[htb]
  \begin{center}
    \caption{Predicted gamma}
   \begin{tabular}{|c||c|c|c|c|} \hline
     x & Training & $\nu_d=\nu_g=0.0$ & $\nu_d=\nu_g=0.01$ & $\nu_d=0.05, \nu_g=0.02$ \\ \hline \hline
     93	&0.0371	&0.0331	&0.0331	&0.0330	\\
     95	&0.0454	&0.0382	&0.0383	&0.0385	\\
     97	&0.0512	&0.0427	&0.0429	&0.0433	\\
    100	&0.0532	&0.0469	&0.0472	&0.0482	\\
    103	&0.0470	&0.0460	&0.0465	&0.0481	\\
    105	&0.0400	&0.0418	&0.0424	&0.0444	\\
    107	&0.0320	&0.0345	&0.0351	&0.0374	\\ \hline
    \end{tabular}
\end{center}
\end{table}

\end{document}